\documentclass{ws-p10x7}

\begin{document}

\title{Radiative $B$ Decays at CLEO}

\author{T. E. Coan}

\address{Physics Department, Southern Methodist University,\\
 Dallas, TX 75275, USA\\E-mail: coan@mail.physics.smu.edu}

\twocolumn[\maketitle\abstract{

We report on the status of a variety of radiative $B$ decays studied
by the CLEO detector with $9.7\times 10^6$ $B\bar{B}$ pairs.}]

\section{Introduction}

Flavor changing neutral currents (FCNC) are well known to be forbidden
at tree level in the Standard Model (SM). At higher order, however,
loop diagrams (box and penguin diagrams) can generate effective flavor
changing neutral currents, i.e., $b\rightarrow s$ and $b\rightarrow d$
transitions. The rates for such transitions in the SM are functions of
the top quark mass as well as the masses of the $W$ and and $Z$ gauge
bosons. Additionally, these rates are also sensitive to the exchange
of heavy non-SM particles such as charged Higgs. Deviations from SM
rates for $b\rightarrow s$ and $b\rightarrow d$ transitions are then a
signature of physics beyond the SM. Hence, measurements of
$b\rightarrow s$ and $b\rightarrow d$ transitions are an effective low
energy probe of physics at a much higher energy scale. This report
emphasizes studies of reactions mediated by electromagnetic penguin
diagrams.

\section{General Experimental Strategy}

All data is taken at a symmetric $e^+e^-$ collider in the vicinity of
the $\Upsilon(4S)$ resonance, just above threshold for $B\bar{B}$
production so that $B$ mesons are produced nearly at rest.
Approximately two-thirds of the data is taken at the resonance and
one-third is taken slightly below the resonance to study continuum and
to perform background subtraction. The total luminosity, summed over
on and off resonance data, is $14.0\,{\rm fb}^{-1}$, corresponding to
$9.7\times 10^6$ $B\bar{B}$ pairs. 

Two observables are particularly
useful for reconstructing $B$ candidates. The first is the difference
in energy between a $B$ candidate and the beam energy, $\Delta E = E_B
- E_{beam}$. The second is the beam-constrained $B$-mass, $M_B=
\sqrt{E^2_{beam} - p^2_B}$, where $p_B$ is the momentum of the $B$
candidate. Additionally, the shape of spherical $B\bar{B}$ events is
used to distinguish them from jet-like continuum events.

\section{Exclusive Modes}

\begin{figure}
\epsfxsize150pt
\figurebox{150pt}{150pt}{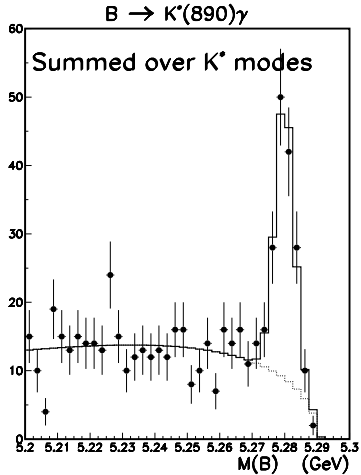}
\caption{The beam-constrained $B$ mass distribution for $B\rightarrow
K^{\ast}\gamma$ summed over the four $K^{\ast}$ modes discussed in the
text.}
\label{fig:kstar}
\end{figure}

CLEO first observed the exclusive mode $B\rightarrow K^{\ast}\gamma$
and continues to collect statistics. The $K^{\ast}$ always decays to
$K\pi$ and CLEO reconstructs four modes ($K^{\pm}\pi^{\pm}, K^0_{\rm
S}, K^{\pm}\pi^0, K_{\rm S}\pi^0$) to be consistent with the
$K^{\ast}(890)$ resonance. No other $K^{\ast}$ resonances lie in this
mass range.  The $K\pi$ system and a hard photon are required to be
consistent with the $B$ mass using the variables $\Delta E$ and
$M_B$. The major background is from continuum ($e^+e^- \rightarrow
q\bar{q}$ with $q = u, d,c,s$) events accompanied by a hard photon
from initial state radiation or events of the type $e^+e^- \rightarrow
(\pi^0,\eta)X$ with $\pi^0,\eta \rightarrow \gamma\gamma$. Both
backgrounds are suppressed with appropriate events shape and
$\pi^0,\eta$ vetoes.  Summed over all $K^{\ast}$ modes, the yield
versus $M_B$ distribution is shown in figure~\ref{fig:kstar}.

Separated into neutral and charged $B$ modes, the measured branching fractions are:
${\cal{B}}(B^+\rightarrow K^{\ast +}\gamma) =(4.5 \pm 0.7 \pm0.3)\times 10^{-5}$ and
${\cal{B}}(B^0\rightarrow K^{\ast 0}\gamma) =(3.8 \pm 0.9 \pm 0.3)\times 10^{-5}$.

CLEO's large sample of $B\rightarrow K^{\ast}\gamma$ decays allows a
search for direct CP violation by measuring the fractional CP
asymmetry parameter $A_{CP} = ({\cal{B}}(b) - {\cal{B}}(\bar{b})/
({\cal{B}}(b) + {\cal{B}}(\bar{b})$, where ${\cal{B}}(b)$ is the branching
fraction of $B^-$ and $\bar{B}^0$ into $K^{\ast}\gamma$ and
${\cal{B}}(\bar{b})$ is the branching fraction of $B^+$ and ${B}^0$ into
$K^{\ast}\gamma$. Here, self-tagging $K^{\ast}$ decay modes are used
to produce the result, summed over charged and neutral $B$ decay
modes, $A_{CP} =+0.08\pm0.13\pm0.03$.

CLEO has searched for resonances heavier than $K^{\ast}(890)$, the
$K_2^{\ast}(1430)$ and the $K^{\ast}(1410)$, each of which has an
appreciable branching fraction to the final state $K\pi$:
${\cal{B}}(K_2^{\ast}(1430)\rightarrow K\pi) =50\pm 1\%$ and
${\cal{B}}(K^{\ast}(1410)\rightarrow K\pi) =7\pm 1\%$. These modes can
be distinguished because of their different helicity distributions and
decay widths. Fitting the $M_B$ distribution for both decay modes
yields the branching fraction results: ${\cal{B}}(B\rightarrow
K_2^{\ast}(1430)\gamma) = (1.7\pm0.6\pm0.1)\times 10^{-5}$ and
${\cal{B}}(B\rightarrow K^{\ast}(1410)\gamma)< 12.7\times 10^{-5}$ at
90\% CL.  

\begin{figure}
\epsfxsize170pt
\figurebox{170pt}{170pt}{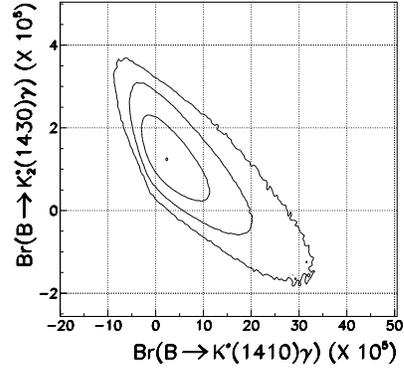}
\caption{Likelihood contours for the simultaneous fit of $B\rightarrow
K_2^{\ast}(1430)\gamma$ and $B\rightarrow K^*(1410)\gamma$ rates. The
central point shows the location of the maximum likelihood. The 1, 2,
and 3 standard deviations from this maximum are indicated by the
contours.}
\label{fig:contours}
\end{figure}

The likelihood contours for the fit are shown in
figure~\ref{fig:contours}. These results imply that $R\equiv
{\cal{B}}(B\rightarrow K^{\ast}(1430)\gamma)/ {\cal{B}}(B\rightarrow
K^{\ast}(890)\gamma ) = 0.4\pm 0.1$, favoring models that use
relativistic form factors\cite{veseli} to predict these rates and
disfavoring those with non-relativistic form factors\cite{ali}.

\begin{figure}
\epsfxsize200pt
\figurebox{180pt}{180pt}{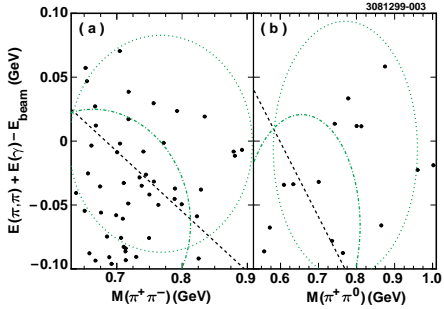}
\caption{The $\Delta E$ vs. $M(\pi\pi)$ distribution for a)
$B^0\rightarrow \rho^0\gamma$ and for b) $B^+\rightarrow \rho^+\gamma$
candidates.  Dots above the slanted line have passed all cuts. The
central dotted ovals are the limits that contain 90\% of the
$B\rightarrow \rho\gamma$ candidates. The partial ovals are the limits
that contain 90\% of the background $B\rightarrow K^{\ast}\gamma$
events.}
\label{fig:bdg}
\end{figure}

CLEO has also searched for $b\rightarrow d \gamma$ transitions of the
form $B\rightarrow\rho\gamma$ and $B\rightarrow\omega\gamma$ which are
a means to limit the CKM ratio $|V_{td}/V_{ts}|$ through the ratio of
branching fractions $R\equiv
{\cal{B}}(B\rightarrow\rho(\omega)\gamma)/ {\cal{B}}(B\rightarrow
K^{\ast}\gamma)=\xi|V_{td}/V_{ts}|^2$, where $\xi$ is the ratio of
$B\rightarrow\rho$ and $B\rightarrow K^{\ast}\gamma$ form factors and
lies in the range 0.6--0.9.  The $\Delta E$ vs. $M(\pi\pi)$
distributions for $B^0\rightarrow\rho^0\gamma$ and
$B^+\rightarrow\rho^+\gamma$ candidates are shown in figure
~\ref{fig:bdg}. The branching fraction limits are
${\cal{B}}(B^0\rightarrow \rho^0\gamma)< 1.7\times 10^{-5}$ and
${\cal{B}}(B^+\rightarrow \rho^+\gamma)< 1.3\times 10^{-5}$ at the
90\% CL. This corresponds to $R< 0.32$ at the 90\%CL.  For the choice
$\xi=0.6$, this implies $|V_{td}/V_{ts}|<0.72$ at 90\% CL. The
search for $B\rightarrow\omega\gamma$ yields the limit
${\cal{B}}(B^0\rightarrow \omega\gamma)< 0.92\times 10^{-5}$ 
at the 90\% CL.

\section{Inclusive Modes}

CLEO has measured the inclusive branching fraction
${\cal{B}}(b\rightarrow s \gamma)$ by determining the hard photons
energy spectrum and then performing an ON/OFF resonance subtraction.
Backgrounds for this analysis are from continuum with initial state
radiation and from continuum with a high energy $\pi^0$, $\eta$, or
$\omega$, where one of the daughter photons escapes detection.  The
ON-resonance background is suppressed by two methods. The first method
uses a neural net (NN) technique based on event shape variables to
separate $B\bar{B}$ events from non-$B\bar{B}$ events. The second
technique uses a pseudo-reconstruction method and a second NN. The end
result is that events are weighted by a NN output. Afterwards,
OFF-resonance data is subtracted.  Photon energies $E_{\gamma}$
between $2.1\,{\rm GeV} < E_{\gamma} < 2.7\,{\rm GeV}$ are used for
the final result. Using only $3.1\,{\rm fb^{-1}}$ of ON-resonance
data, CLEO measures ${\cal{B}}(b\rightarrow s \gamma) =
(3.15\pm0.35\pm0.32\pm0.26) \times 10^{-4}$, consistent with SM
calculations.



CLEO has searched for direct CP violation in $b \rightarrow s\gamma$
decays for the full data sample of $9.7\times10^6$ $B\bar{B}$ events
by constraining the fractional CP violation parameter ${\rm A_{CP}}$.
Although in the SM, ${\rm A_{CP}}$ is expected to be less than 1\%,
some non-SM physics scenarios\cite{kagan}${}^{,}$\cite{aoki} permit ${\rm
A_{CP}}< 10-40\%$. Events are flavor tagged using either the charge of
a high momentum lepton from the ``other'' $B$ or by using a
pseudo-reconstruction technique similar to the one used in the
inclusive $b\rightarrow s\gamma$ analysis. The typical mis-tag rate is
10\%.  CLEO finds $-0.22<{\rm A_{CP}}< +0.09$ at the 90\% CL.

\section{Non-penguin Radiative Decays}

\begin{figure}
\epsfxsize170pt
\figurebox{170pt}{170pt}{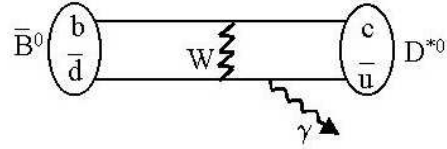}
\caption{Feynman diagram for the decay $\bar{B}^0 \rightarrow D^{\ast 0}\gamma$.}
\label{fig:savinov}
\end{figure}

Finally, CLEO has searched for radiative $B$ decays not mediated by
electromagnetic diagrams, as shown in figure~\ref{fig:savinov}. In the
SM such decays are expected to be small but their observation would
permit the importance of $W$ exchange diagrams in B decays to be
gauged. Here, the $ D^{\ast 0}$ decays by $\pi^0$ or $\gamma$ emission
and the $D^0$ is reconstructed in the $K^-\pi^+, K^-\pi^+\pi^0,
K^-\pi^+\pi^-\pi^+$ final states. No events are found in $M_B -\Delta E$ space
which leads to the branching fraction limit\cite{savinov} ${\cal{B}}(\bar{B}^0 \rightarrow 
D^{\ast 0}\gamma) < 5\times 10^{-5}$ at the 90\% CL.

\section*{Acknowledgments}
The kind assistance of T. Skwarnicki is noted.

\end{document}